# Exploring the microstructural origins of conductivity and hysteresis in metal halide perovskites via active learning driven automated scanning probe microscopy

Yongtao Liu,[1,a] Jonghee Yang,[2] Rama K. Vasudevan,[1] Kyle P. Kelley,[1] Maxim Ziatdinov,[1,3] Sergei V. Kalinin,[2,b] and Mahshid Ahmadi[2,c]

[1] Center for Nanophase Materials Sciences, Oak Ridge National Laboratory, Oak Ridge, TN, 37830, USA

[2] Department of Materials Science and Engineering, University of Tennessee, Knoxville, TN, 37996, USA

[3] Computational Sciences and Engineering Division, Oak Ridge National Laboratory, Oak Ridge, TN 37830, USA

Electronic transport and hysteresis in metal halide perovskites (MHPs) are key to the applications in photovoltaics, light emitting devices, and light and chemical sensors. These phenomena are strongly affected by the materials microstructure including grain boundaries, ferroic domain walls, and secondary phase inclusions. Here, we demonstrate an active machine learning framework for "driving" an automated scanning probe microscope (SPM) to discover the microstructures responsible for specific aspects of transport behavior in MHPs. In our setup, the microscope can discover the microstructural elements that maximize the onset of conduction, hysteresis, or any other characteristic that can be derived from a set of current-voltage spectra. This approach opens new opportunities for exploring the origins of materials functionality in complex materials by SPM and can be integrated with other characterization techniques either before (prior knowledge) or after (identification of locations of interest for detail studies) functional probing.

[a] liuy3@ornl.gov
[b] sergei2@utk.edu
[c] mahmadi3@utk.edu

Metal halide perovskites (MHPs) have emerged as the most promising class of materials for the next generation of photovoltaics due to their high photovoltaic efficiency and potential for low-cost large-scale manufacturing.[1, 2] In addition to single junction MHP solar cells, MHP allows for bandgap engineering via mixing halides, making MHPs suitable for tandem solar cells when stacking with Si or other thin film photovoltaics. However, the current issue for commercializing MHP solar cells is their stability and scalability.[3-6]

Both stability and scalability are ultimately related to the microstructure of the film and its evolution during deposition and annealing. The formation of the microstructure starts at the deposition stages, and is controlled by nucleation and subsequent microstructure evolution during annealing and processing.[7] The resultant microstructures are often inhomogeneous and contain grain boundaries, topological defects such as domain walls, and secondary phase inclusions.[8] All these elements can affect the carrier accumulation and electron and hole transport.[6] Additionally, these elements can strongly affect the ionic dynamics during operation and can be susceptible to environmental stresses, giving rise to light-, bias- and ambient gas-induced instabilities in these materials.[8, 9]

These considerations have stimulated studies of the chemical and physical microstructure of the MHP films using a variety of microscopic imaging and chemical characterization techniques.[10-18] Multimodal microscopy measurements revealed the coupled chemical, strain, and ferroic behavior,[19] which affects the optoelectronic response,[20] implying that it can play a role in charge carrier transport and generation and ultimately affect photovoltaic performance. It was also reported that local chemical disorder results in electronic disorder and charge carrier transport.[21] Local phase impurities, depending on the thin film composition and processing method, act as trap sites, which induces structural change and photochemical degradation under illumination.[8] Using time-resolved secondary ion mass spectrometry to observe real-time chemical distribution in MHP film, researchers revealed chemical redistribution and associated degradation induced by external fields including light and electric fields.[22-26] Ferroic twin domains in MHPs have been found to correlate with both crystallographic difference and chemical heterogeneity.[19, 27, 28] These nanoscale domains lead to photoluminescence intensity differences,[20, 29] which was attributed to the difference in local charge carrier concentration. On one hand, the domain walls are found to be benign to charger carrier activity;[29] on the other hand, the charge carrier mobility is shown to be different along the domain wall and perpendicular to the domain wall direction.[15] In addition, sub-grain boundaries can act as energetic barriers to block charge carrier transports.[30] Moreover, significant work has revealed that grain boundaries can serve as trap sites, recombination centers, ion migration highways, etc., and ultimately affects the efficiency and stability of MHP photovoltaics.[31]

This extensive variability of microstructural elements on length scales spanning nanometer scale phase separated regions and potentially larger scale defects that can emerge during industry-scale manufacturing necessitates investigation of local functionalities to build structure-property relationships. Direct information on the local transport properties can be obtained by conductive AFM (cAFM). In this technique, the AFM tip plays a role of nano electrode that establishes the contact with individual regions of the surface. The measured current-voltage (IV) curve contains information on local electronic and ionic conductivity. The applications of more complex waveforms such as up-down sweeps and first-order reversal curve (FORC) IVs additionally allows to identify the hysteresis (typically related to ionic motion or electrochemical processes).[25, 32] While cAFM is affected by factors such as topography, etc., the collected information revealing local conductivity with a high spatial resolution is directly relevant to device operation.

However, measuring current-voltage characteristics everywhere across the sample surface is highly non-optimal. Measurements over dense grid can be time consuming, associated with accumulated damage to the AFM tip and surface, and are not guaranteed to sample the object of interest such as GBs if the spatial density of the latter is low. Previously, we have introduced a direct computer vision-based method where we use pretrained supervised network[33-35] to identify objects of interest and explore their transport properties[36]. Here, we implement the solution of the opposite problem–namely discover what region of the surface and microstructural element are responsible for a specific aspect of transport behavior.

Recently, we developed a deep kernel (active) learning (DK(A)L)[37-39] approach, which is a machine learning workflow that can actively learn the correlation between structure image data and functional (e.g., spectroscopy) data during an operating experiment, and subsequently makes decisions on next measurement locations based on (continuously updated) learned relationship. Here, the image includes structural features such as grain or grain boundary (GB), and the spectroscopy data (e.g. IV curve) which encodes the physical properties. We designed a workflow to implement the DKL on an operating AFM, allowing us to explore the materials in an automated manner without human intervention. In this work, we implement this approach in cAFM to explore the conductive properties in MHP thin films. We use the AFM topography that shows the grain and GB as the structure image data for DKL, and perform IV measurements as the spectroscopy data. Then, the DKL actively learns the correlation between topography and IV curves. The DKL-cAFM results indicate that GB junction points show interesting phenomena, while previous work only focused on the difference between grain and GB.

As a model system, a mixed formamidinium (FA) and cesium (Cs) cation (FACs) MHP (FACs-MHP) is selected for this study. This FACs-MHP is one of the state-of-the-art MHPs, which reports show excellent stability and a PCE of 22.7% for FACs-MHPs solar cells.[40, 41] Nevertheless, realizing a completely phase-pure FACs-MHP is challenging due to the relatively high Cs concentration (<20%), causing local microstructural inhomogeneities.[40, 42, 43] Thus, FACs-MHPs are the ideal platform for the microstructural explorations using DKL-cAFM.

Traditionally, cAFM is operated in an image or spectroscopy mode to study MHP thin films, as shown in Figure 1. Here, the image measurement indicates the conductivity of the film under a static condition, while the spectroscopy measurement shows IV results from certain locations determined by human operator. However, for traditional cAFM, spectroscopy locations must be manually selected and triggered by human operators; thus, only a relatively limited number of points can be selected, and the selected locations are based on operators' interest.

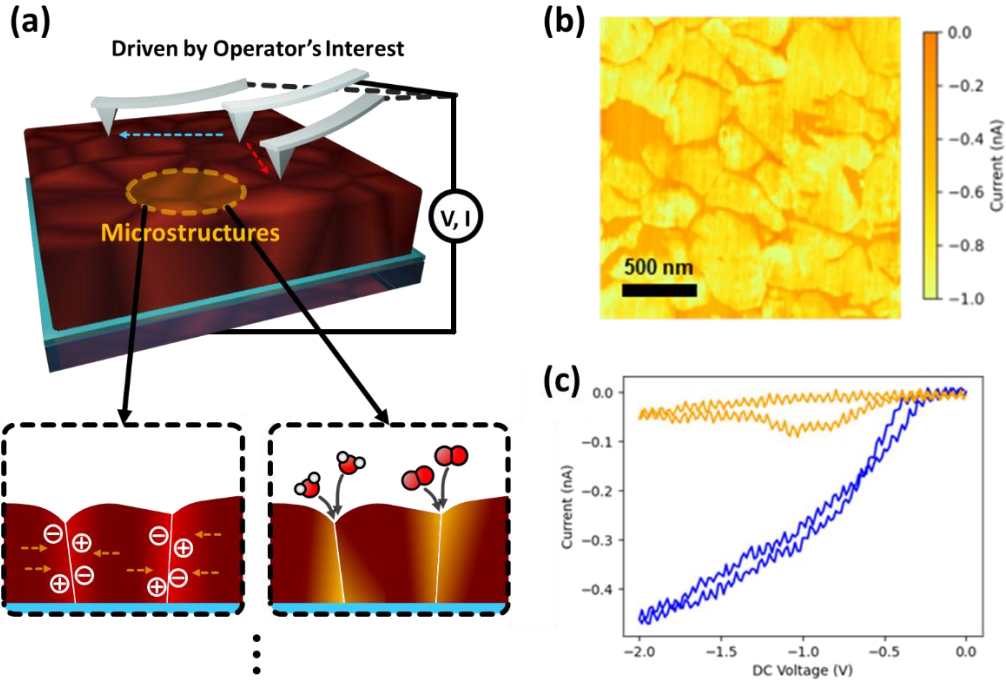

**Figure 1.** Classical cAFM studies of MHP. (a) Schematic representation of cAFM measurement exploring microstructural effects (e.g., grain boundaries) on the charge transport in MHPs. This includes local compositional inhomogeneity, charge accumulation and phase transformation via environmental stresses (e.g. $H_2O$ or $O_2$ penetration). (b) In image mode, cAFM obtains current maps by applying a static voltage -1.5 V. (c) In spectroscopy mode, human operator can manually select spectroscopy measurement locations, and subsequently trigger spectroscopy measurements.

Next, we employ the recently developed deep kernel learning (DKL) method to explore the relationship between IV spectroscopy and surface morphology in an unbiased manner. In DKL analysis, the image (e.g. topography) data showing microstructure is the input and the spectroscopy (e.g. IV) showing the physical property is the output, as shown in Figure 2a. The DKL includes a neural network (NN) and a Gaussian process (GP) layer, where the NN featurizes the structural data into a small number of latent variables and the GP operates over this latent space to analyze its relationship with spectroscopy data. The workflow schematic of DKL operating in an autonomous setting on the microscope is shown in Figure 2b. Here, a raster scan of the topography is first acquired as the structural image followed by an IV spectroscopy measurement that is performed at a random or pre-defined location. The DKL is then trained with this IV spectroscopy data and the corresponding image structure, and subsequently the trained DKL will predict the spectroscopy property at other unmeasured locations. Based on the DKL prediction, an acquisition function, i.e. maximum uncertainty, is used to derive the next measurement location. Then, the workflow transfers this location to the microscope and drives the microscope to perform next IV measurement. The above process will be repeated to actively explore the morphology-IV relationship. It should be noted that the DKL model (including both the neural net and GP) is updated as new spectroscopy data is acquired, i.e., it is in an active learning setting.

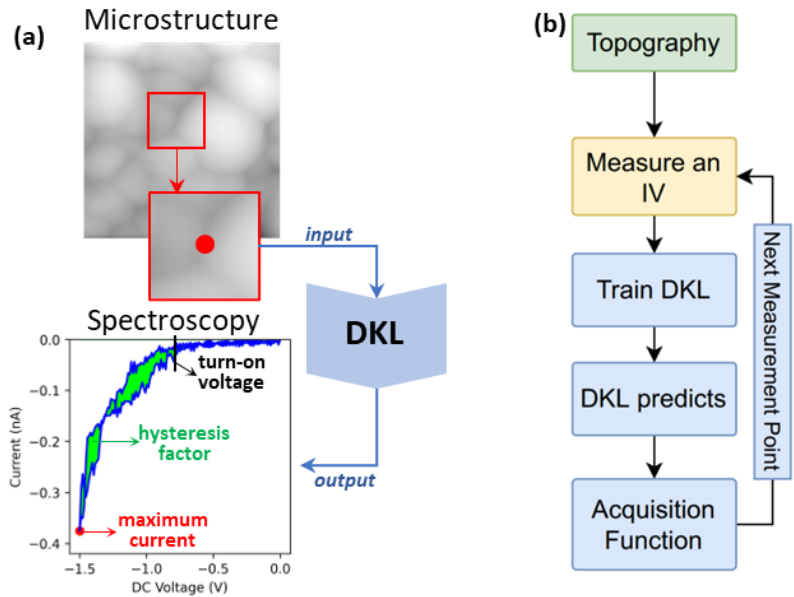

**Figure 2.** DKL-cAFM. (a) the input of DKL is an image data showing the microstructure (e.g. grain or grain boundary), the output of DKL is spectroscopic data including functionality of interest, DKL analyzes the correlation between input and output. (b) the workflow of DKL-cAFM.

We define several physical descriptors representing the property of IV curve as target properties in the DKL-cAFM exploration. As shown in the spectroscopy IV example in Figure 2a, we define several descriptors including the voltage where the current starts increasing as the turn on voltage; the difference between forward and reverse IVs as the hysteresis factor; and the maximum current. These descriptors are extracted from real-time IV curves and used as the target property for DKL exploration. It is well-known that light illumination changes strain disorder, ion migration activation energy, phase segregation, etc. in MHPs and consequently alters MHP properties. As such, we performed DKL-cAFM exploration in both dark and light conditions. In this work, we used a built-in LED light in the microscope to illuminate the MHP film, the light intensity is roughly equivalent to 0.1 sun.

Shown in Figure 3 is the results of DKL-cAFM exploration over the whole topography region. Here, topography is used as the structure image and three physical descriptors—turn on voltage, hysteresis factor, and maximum current—are used as target properties to guide DKL-cAFM exploration. Also, each exploration is performed under dark and light conditions. Overall, six DKL-cAFM explorations are performed. The topography image size is 128 x 128, totally 16384 pixels (locations); in DKL-cAFM measurement, we performed 200 exploration steps that obtained IV from 200 locations, which is ~1.2% of total locations. In the DKL-cAFM measurement, the DKL analysis is performed on a GPU server (Nvidia DGX-2). A real-time communication between AFM workstation and remote GPU server is built via socket programming, so that the experiment data stream is sent to the GPU server and DKL analysis result is sent back to AFM workstation for the subsequent measurement. After 200 steps measurement, DKL is able to offer a prediction map of the target property of the whole region and an associated uncertainty map showing the variance in the prediction at each location.

In Figure 3, the filled circles in the topography image are the IV measurement locations decided by DKL, where the blue spots indicate the measurement locations at the beginning and the red spots indicate the measurement locations at the end of the measurement, i.e., the color indicates the measurement time step. A universal phenomenon we can observe from all six measurements (note that six measurements are independent) is that the measurement locations tend to concentrate around GBs regardless of target property or light condition. To the MHP community, this is not a surprising result as GBs are known to exhibit multiple functional properties, such as acting as defect centers, traps, ion migration highway, etc. This result is a good indicator that DKL exploration is based on physics. A close look at the results in Figure 3 leads to a discovery that the measurement locations not only concentrated near GBs but also the GB junction points (the cross point of several GBs). To our knowledge, though previous work has largely focused on investigating the difference between grain and GBs, the GB junction points were rarely investigated. This result indicates that GB junction points potentially exhibit interesting functionality, which is a discovery by DKL active learning with human intervention. In this regard, we direct readers to our recent study[36] with supervised machine learning method to systematically investigate GBs, in which we clarified that GB junction points are insulating under dark and conductive under light. Such phenomenon is the related to spatial chemical composition variance. In addition, in the DKL prediction, spatial variation of each target property is observed (Figure 3). This variation is mostly associated with grain structure, which may relate to intrinsic defects distribution, ion migration, or local phase impurities.

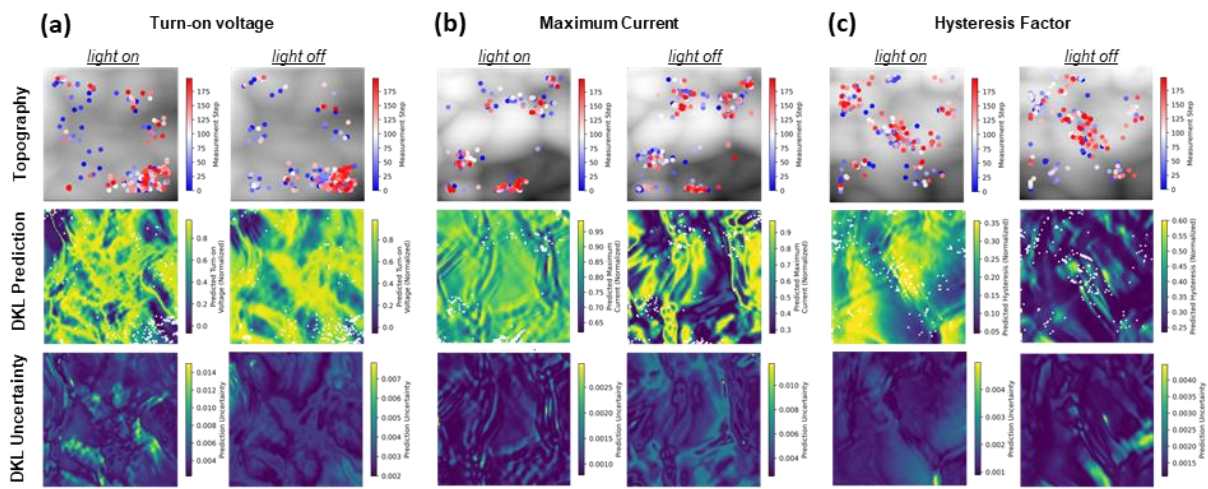

**Figure 3.** DKL-cAFM results, where the featurized topography data is used as DKL inputs and the properties are extracted from IV curves (e.g. turn-on voltage, maximum current, hysteresis factor) used as the target property. For each target property, the measurement is performed in both light-on and light-off conditions. The DKL exploration locations are shown in the topography image, where the blue filled circles indicate the exploration location at the beginning and the red filled circles indicate the exploration location at the end. (a) target property is turn-on voltage; (b) target property is maximum current; (c) target property is hysteresis factor.

Since the DKL exploration over the whole region indicates GBs to be interesting, a further exploration focusing on GBs is necessitated. A pre-requirement to explore over GBs is identifying GB locations, which can be done by the ResHedNet model developed recently, which is a modified version of holistically nested edge detector[44] augmented with residual connections.[34] To improve the robustness of ResHedNet predictions on real-time data streams, an ensemble of ResHedNet models—each model is trained independently—is used in the real-time experiment. Compared to traditional edge detectors as well as a single ResHedNet, this ensembled-ResHedNet overcomes the out-of-distribution effect in real-time experiments and hence more robust. We trained the ensemble-ResHedNet to identify the locations of the GBs, then we extracted topography patches from the GBs locations and made these patches as exploration space (i.e. now the exploration space is only GBs). In this instance, in the DKL-cAFM experiment, the DKL only explores the morphology-IV relationship at GBs.

In the same manner as the previous measurements, the topography is used as the structural image and three physical descriptors are used as the target property to guide DKL-cAFM measurement at GBs. The results are shown in Figure 4. Again, the exploration locations decided by DKL show a tendency to concentrate around GB junction points regardless of whichever target property is used. In addition, an obvious observation in Figure 4 is that the DKL procedure tends to explore some GBs but not others—that is, exploration locations mostly distributed in certain GBs. This can be understood as the internal variation of GBs, i.e., different GBs may exhibit different functionalities.

The DKL prediction of turn-on voltage draw a map showing complex features around the GBs under light that show higher values under dark. The predicted maximum current map show the higher levels under light across the surface, attributed to the carrier photogeneration in the MHP matrix. The predicted hysteresis factor map also exhibits local complexities, whereas smaller hysteresis factors are predicted around the GBs under light and dark conditions. Together, the resulting DKL predicted maps reveal local complexities in the MHP matrix directly visualizing the morphology-charge transport relationship, which have been undermined based on classical operator's interests.

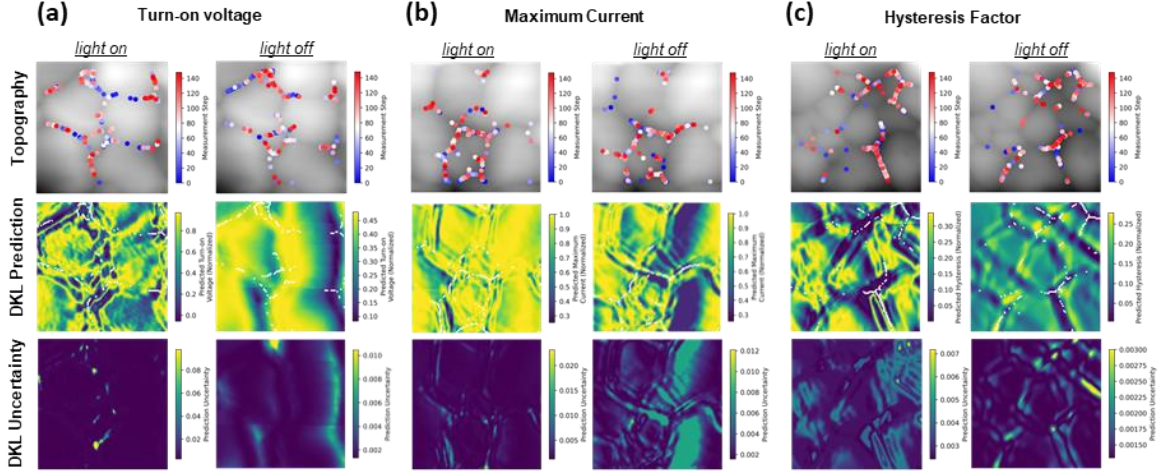

**Figure 4.** Res-DKL-cAFM results, where the topography is used as the structural image and the properties are encoded in IV curves (e.g. turn-on voltage, maximum current, hysteresis factor), which are used as the targets. For each target property, the measurement is performed in both light-on and light-off conditions. The ResHedNet is trained to find GBs, then the topography image patches are extracted from GBs as the DKL exploration space, as such DKL only explores GBs. The DKL exploration locations are shown in the topography image, where the blue filled circles indicate the exploration location at the beginning and the red filled circles indicate the exploration location at the end. (a) target property is turn-on voltage; (b) target property is maximum current; (c) target property is hysteresis factor.

To obtain further insight into the DKL results, we introduce the approach for the post-experiment data analytics that we refer to as post-acquisition forensics. In this case, after the DKL experiment we have access to (1) the series of local image patches acquired by the DKL algorithm during the experiment, $P_i$, where $i = 1, 2, .., n$ is the image patch at step $i$. During the DKL experiment, the DKL algorithm sequentially explored images $P_i$ trying to optimize the chosen characteristic of the IV curve in its center; (2) The full set of image patches can be formed from the original AFM image; (3) the IV curves $IV_i$ measured at the center of image patch $i$.

As a first step of analysis, we explore the spatial variability of the sampled image patches compared to the full set of possible image patches. To accomplish this, we use a rotationally invariant variational autoencoder (rVAE) approach to discover the intrinsic factors of variability within the data, as described in detail in our previous work.[34, 45, 46] Briefly, rVAE encodes the information in the image patch $P_i$ into the latent vector L. Here for the ease of representation we choose L to be 2D, L = (L$_1$, L$_2$). The important aspect of the rVAE approach is that it allows us to disentangle the representation of the data, meaning, to discover the factors of variability in the data set and align them with specific latent variables. With rVAE, the collection of patches can be represented as a distribution in the 2D space as shown in Figure 5. Here, we illustrate the latent distribution of the full data set via the corresponding kernel density estimate (KDE) and illustrate the distribution of the points corresponding to the sampled patches P$_i$ via color-coded points. It is clearly visible from this analysis that the experimentally probed density (Figure 5b) is very different from the full density of the system (Figure 5a), so the active learning workflow probes a specific manifold in the full space of the system.

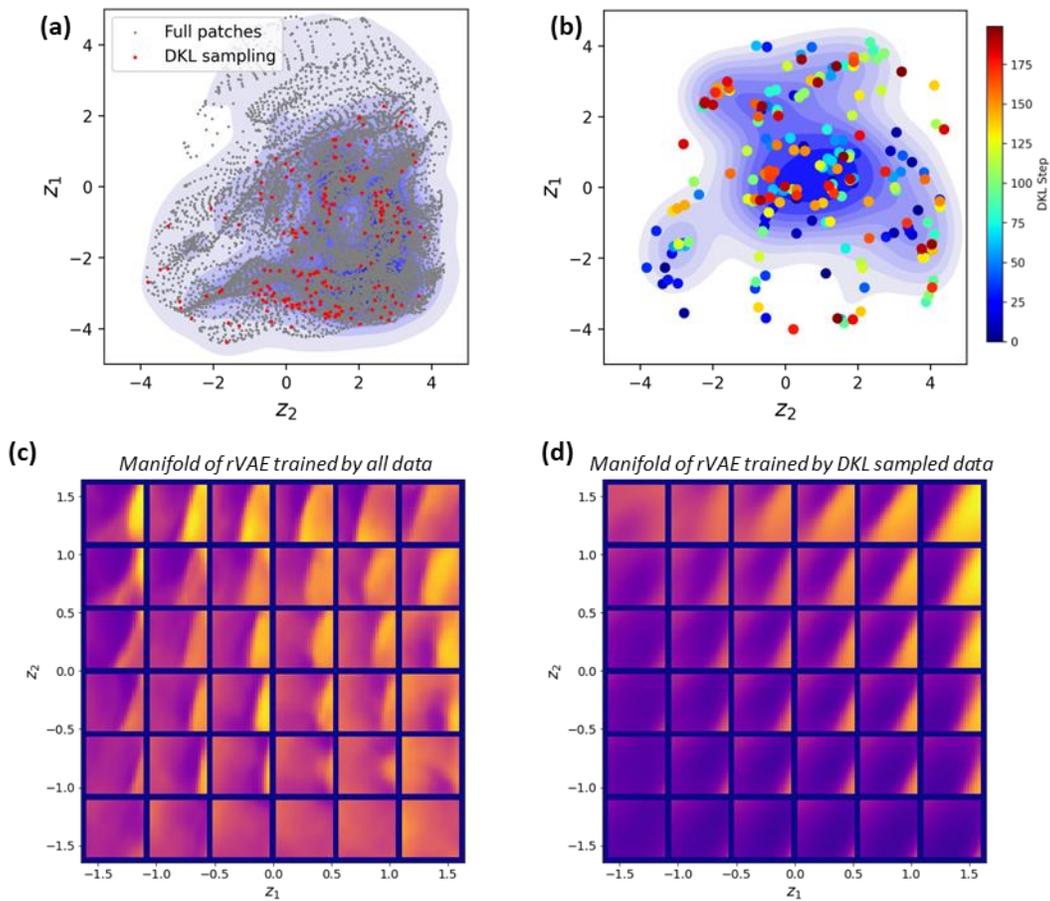

**Figure 5.** rVAE analysis of full data and DKL sampled data. (a), (b), latent distribution of the full data and DKL sampled data, respectively. (c), (d), Latent representation of rVAE trained by full data and DKL sampled data, respectively.

The meaning of the latent variables, i.e. relationship between the latent vector and real space image patch can be illustrated using the latent representation as shown in Figure 5c-d. To construct the latent representation, the rectangular grid of points in the latent space is constructed and each latent vector is decoded to yield the real space image. The latent representation for the full data set is shown in Figure 5c. This shows the key elements of morphology ordered along the latent directions—along the vertical direction, we observe that GB gradually disappear from top to bottom; along the horizontal direction, the GB location changes gradually; in both directions, we observe the change of GB shape, e.g. curvature, as well as a contrast change. In comparison, the latent representation for the image set sampled during the DKL experiment is shown in Figure 5d. Here, the predominate difference among grids is the GB location and contrast. We note that from the material perspective, we also studied the GB behavior in details in a separated work.[36]

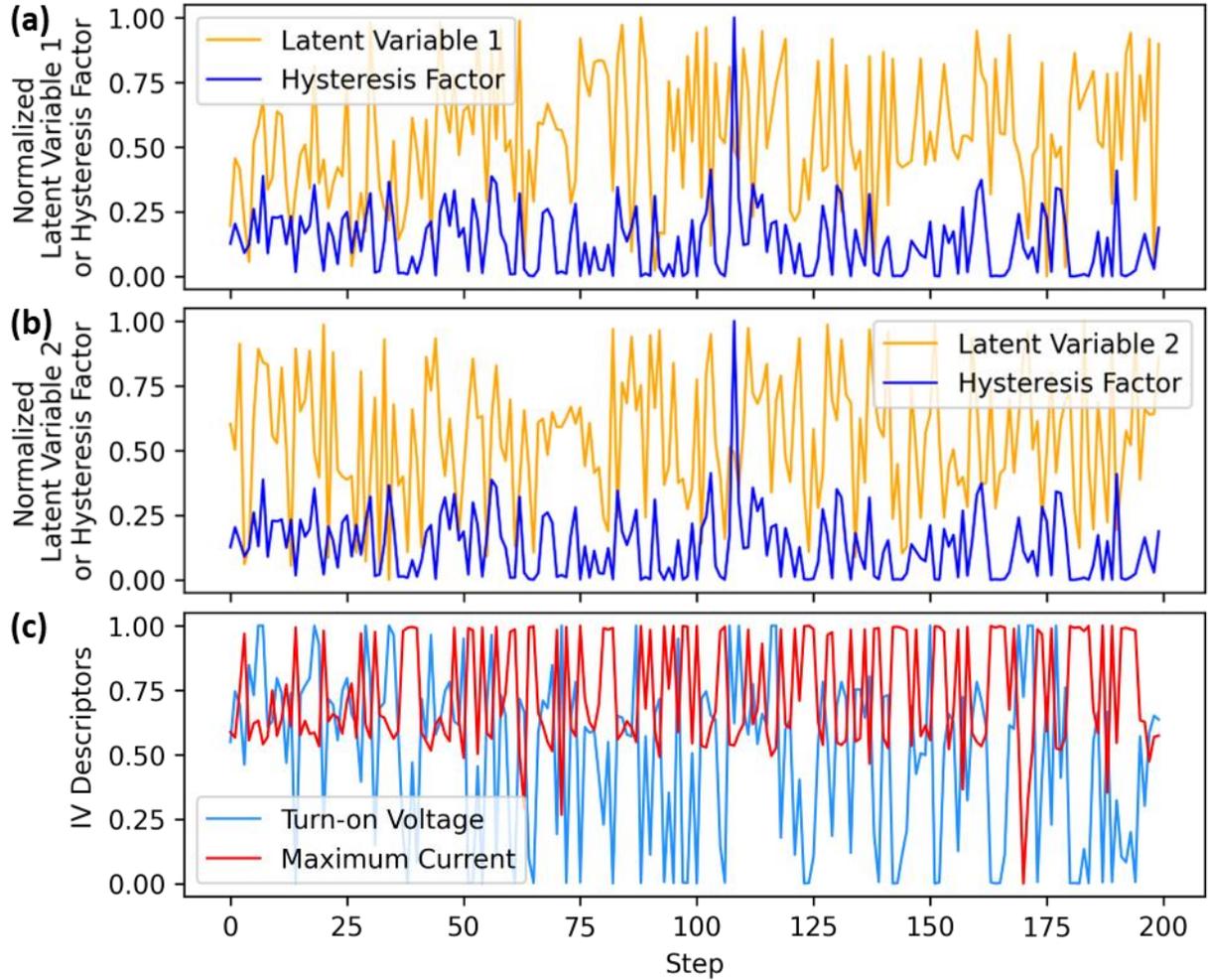

**Figure 6.** Evolution of IV descriptors during DKL experiment and comparison with rVAE latent variables.

With this information in hand, we proceed to analyze the experimental trajectory. Shown in Figure 5b, the color represents experiment steps showing the trajectory of the DKL experiment in the latent space. We further show the evolution of the scalarizer function during the experiment, and compare them with latent variables in Figure 6. Here, this experiment is guided by hysteresis factor, Figure 6a-b show the correlation between hysteresis factor and two latent variables. Figure 6c indicates how turn-on voltage and maximum current evaluate during the experiment.

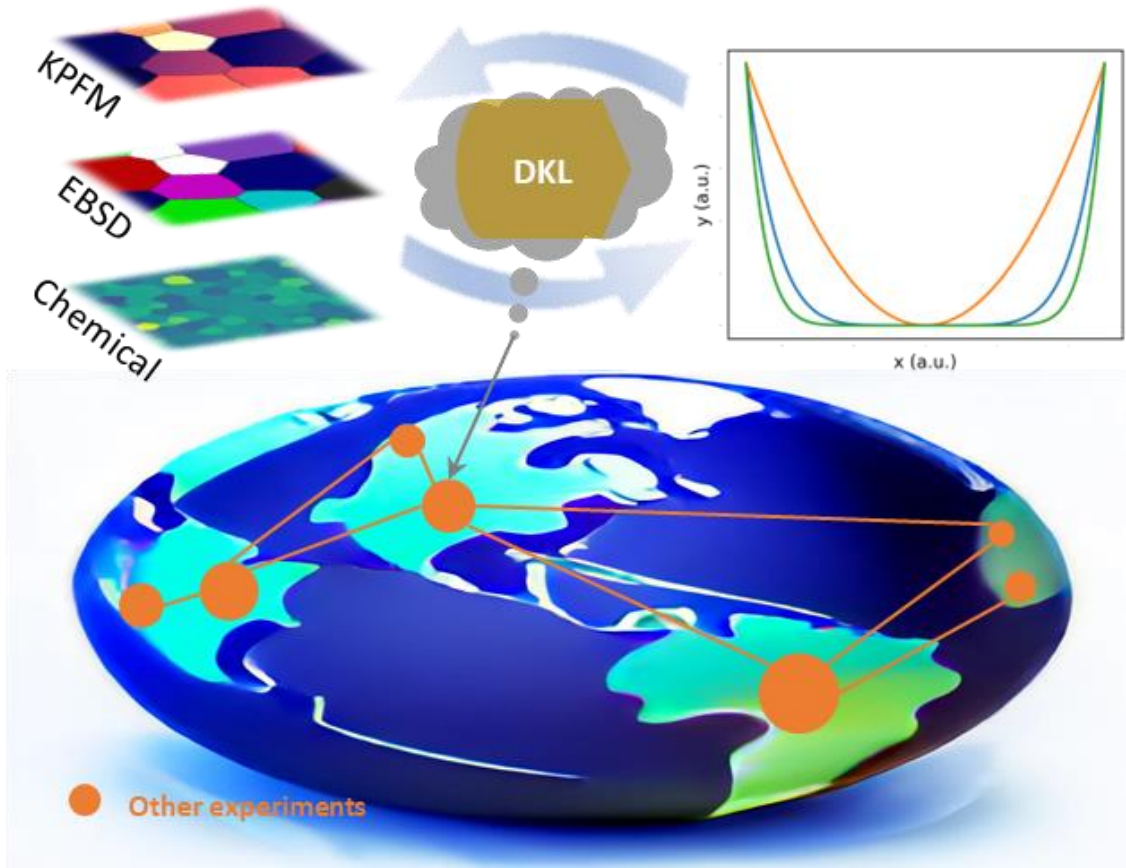

**Figure 7.** Schematic of more general workflow of DKL, where the trained DKL in one measurement can be used in other measurement for exploration as well. The planet picture is generated by DALLE-Mini.[47]

In summary, we implemented a DKL algorithm in cAFM to study the surface morphology-conductivity relationship of a polycrystalline MHP thin film. The approach combines the power of machine learning and AFM, allowing an automated exploration without human bias. The DKL exploration over the whole region including both grain and GBs implies that GBs are of interest, which matches physics that GBs are known to exhibit rich functionality. In addition, the results also show that GBs junction points can be more interesting sites as DKL tend to learn the GB junction points during experiment. We further combined our ResHedNet with DKL that enables DKL focusing on studying GBs instead of the whole region. Interestingly, the results further confirm the GB junction points can exhibit interesting functionality which requires more in-depth study. The exploration of GBs also indicates a variation between GBs. Then, we explored the DKL sampling mechanism via VAE analysis.

In spite of intensive efforts investigating microscale properties of MHP, most work focused on the difference between grain and GBs, the variation between different GBs, as well as the GB junction points, is rarely investigated. This work, by active learning driven cAFM, highlights the role of GB junction points and the variation of GBs in MHP functionality in an automated manner without human bias.

We further note that the DKL approach developed here can be incorporated in more complex machine learning driven exploration pathways. Here, the DKL was realized to establish the relationship between the transport properties and microstructure parametrized via surface topography. However, in the upstream direction (Figure 7), this workflow can be based on the more complex characterization methods including cathodoluminescence, photoluminescence, time-of-flight secondary ion mass spectrometry (ToF-SIMS), electron backscatter diffraction (EBSD), etc. The key requirement here will be either the multimodal imaging when SPM is incorporated in these tools, or creating of the fiducial marks that allow alignment of images acquired from different modalities. Similarly, downstream the region of interest identified via DKL SPM can be explored via destructive methods such as scanning transmission electron microscopy or atom probe tomography to establish the origins of the observed phenomena on the atomic level.

**Methods:**

*MHP synthesis:*

1.0M $Cs_{0.17}FA_{0.83}PbI_3$ precursor solution was prepared by mixing respective volumetric amounts of $FAPbI_3$ and $CsPbI_3$ precursors. A mixture of DMF and DMSO [DMF (v) : DMSO (v) = 5:1] is used as the solvent. . The $Cs_{0.17}FA_{0.83}PbI_3$ precursor was spin-coated on the ITO substrates with 500rpm for10 seconds (ramp 1000 rpm/s) and 4000 rpm for 35 seconds (ramp 2000 rpm/s) to make the thin film. Antisolvent chlorobenzene was added 10 s before the end of spin coating. The films were then annealed at 150 °C for 10 min. The whole processing was performed in $N_2$ glovebox.

*cAFM:*

Conductive AFM measurement were performed with Budget Sensor Multi75E-G Cr/Pt coated AFM probes (~3 N/m) in an Oxford Instrument Asylum Research Cypher microscope. A National Instrument DAQ card is equipped in the Cypher microscope and LabView script is used to apply voltage and acquiring IV data. The automated experiment workflow is designed in a Python Jupyter Notebook with deep learning methods from AtomAI[48].

**Acknowledgements**

This research (DKL) was supported by the Center for Nanophase Materials Sciences (CNMS), which is a US Department of Energy, Office of Science User Facility at Oak Ridge National Laboratory. This effort (implementation in SPM, measurement, data analysis) was primarily supported by the center for 3D Ferroelectric Microelectronics (3DFeM), an Energy Frontier Research Center funded by the U.S. Department of Energy (DOE), Office of Science, Basic Energy Sciences under Award Number DE-SC0021118. J.Y. and M.A. acknowledge support from National Science Foundation (NSF), Award Number No. 2043205.

**Conflict of Interest Statement**

The authors declare no conflict of interest.

**Authors Contribution**

Y.L. and S.V.K. conceived the project. M.Z. developed DKL, ResHedNet, and rVAE. Y.L. implemented DKL in cAFM and generated experimental results. J.Y. and A.M. synthesized MHP film. All authors contributed to discussions and the final manuscript.

**Data Availability Statement**

The method that supports the findings of this study are available at https://github.com/yongtaoliu/MHP_DKL_IV.